\documentclass[10pt,letterpaper]{article}

\usepackage{opex3}
\usepackage{graphicx}
\usepackage{amssymb}

\usepackage{graphicx}
\usepackage{epstopdf}

\begin{document}

\title{Vacuum squeezed light for atomic memories at the $D_2$ cesium line}

\author{Sidney Burks, J\'{e}r\'{e}mie Ortalo, Antonino Chiummo, Xiaojun Jia, Fabrizio Villa, Alberto Bramati, \\Julien Laurat, and Elisabeth Giacobino}
\address{Laboratoire Kastler Brossel,
Universit\'{e} Pierre et Marie Curie, Ecole Normale Sup\'{e}rieure,
CNRS, Case 74, 4 place Jussieu, 75252 Paris Cedex 05, France}
\date{\today}
\email{elg@spectro.jussieu.fr}
\begin{abstract}
We report the experimental generation of squeezed light at 852 nm, locked on the Cesium $D_2$ line. 50$\%$ of noise reduction down to 50 kHz has been obtained with a doubly resonant optical parametric oscillator operating below threshold, using a periodically-poled KTP crystal. This light is directly utilizable with Cesium atomic ensembles for quantum networking applications.
\end{abstract}

\noindent During the last two decades, a great effort has been dedicated to the generation of non-classical states of light in the continuous variable regime. Very recently, 10 dB of noise reduction was obtained with the goal of surpassing the standard quantum limit for sensitive measurements such as gravitational wave detection \cite{Schnabel}.  Driven by the prospect of interfacing light and matter for quantum networking applications\cite{zoller05,cerf,jeff}, ongoing efforts have also focused on the generation of squezeed light at atomic wavelengths and reaching low noise frequencies to be comptatible with bandwidth-limited interfacing protocols. Results have been obtained on the rubidium $D_1$ line\cite{tanimura,hetet} and squeezed light has been recently stored\cite{honda,appel}. Squeezing resonant with the cesium $D_2$ line was demonstrated already a while ago in connection with sub-shot noise spectroscopy experiments \cite{polzik,marin} but not at low-frequency sidebands. More recently, low-frequency squeezing was finally reported at this wavelength \cite{nielsen}, however without giving spectrum  behavior in this range. Here, we give detailed measurements of such squeezing, demsontrating a broadband noise reduction in the low-frequency domain. Furthermore, our setup shows the first usage of a PPKTP crystal for creating squeezed light  at 852nm.  This limits the generally observed losses caused by blue light induced infrared absorption, which are the limiting factor for squeezed light generation.  

\begin{figure}[t!]
\begin{center}
\includegraphics[width=11.5cm]{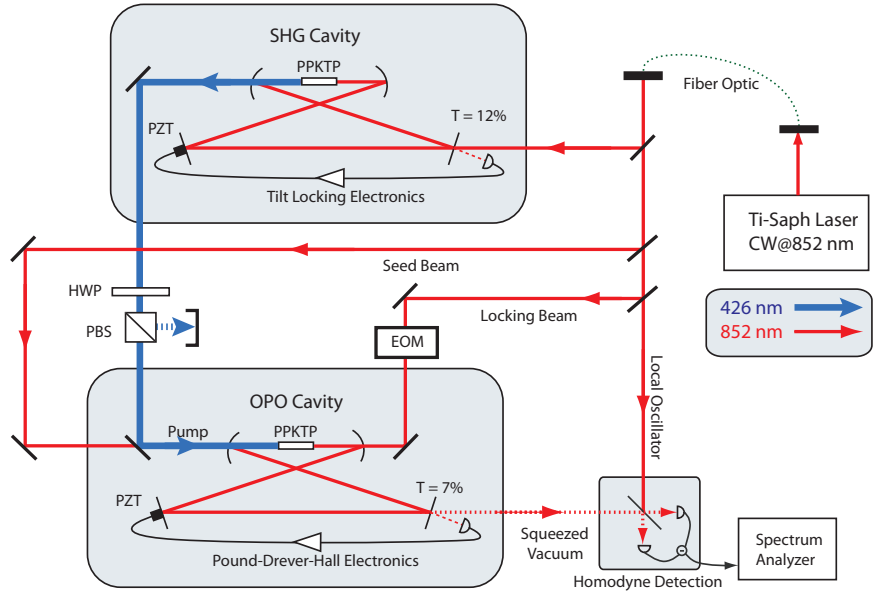}
\end{center}
\caption{Experimental Setup. A Ti:sapphire laser locked on resonance with the Cesium D2 line is frequency doubled. The second harmonic is then used to pump a doubly resonant optical parametric oscillator below threshold.  The seed beam is used for cavity alignment and blocked during measurements. HWP: Half-wave plate. EOM: electro-optic phase modulator. PZT: piezo-electric transducer. PBS: polarizing beam-splitter. 
}\label{setup}
\end{figure}

The experimental setup is sketched in Fig.
\ref{setup}. A continuous-wave
Ti:Sapphire laser (Spectra Physics-\textit{Matisse}) locked on the cesium $D_2$ line is frequency-doubled  in a bow-tie cavity with
a type-I 20 mm long periodically-poled KTP crystal (PPKTP, Raicol Crystals Ltd.) \cite{LKB1}, and locked by tilt-locking\cite{tilt}. By supplying 600 mW of light at 852 nm, we obtain 200 mW of 426 nm cw-light. Higher doubling efficiency can be obtained but with lower stability due to thermal effects. This beam pumps a 550 mm long doubly-resonant (signal and idler) optical parametric oscillator (OPO), based on
a 20 mm long PPKTP crystal. The OPO is locked at resonance using the Pound-Drever-Hall technique \cite{PDH} (20 MHz phase modulation), thanks to a 8 mW additional beam injected through a HR mirror and propagating in the opposite direction of the pump beam. The crystal temperatures are
actively controlled, with residual oscillation of the order of few mK. Both cavities have the same folded-ring design. The crystals are placed between high-reflecting mirrors with a radius of curvature R=100 mm for the OPO, and R=150 mm for the doubler while the other mirrors are flat. The input mirror for the doubler has a transmission of 12$\%$, and the output mirror for the OPO of 7$\%$.  The folding angles are around 10$^{\circ}$, with a cavity length of 55 cm. The waist inside the crystal is around 46 $\mu m$. In this configuration, the OPO threshold is measured to be 90 mW, with a degeneracy temperature at 46.3$^{\circ}C$.  The homodyne detection is based on a pair of balanced high quantum efficiency Si photodiodes (FND-100, quantum efficiency: 90\%) and an Agilent E4411B spectrum analyser. The light from the Ti:Sapphire laser is used after initially being transmitted into a single mode fiber, which improves the matching of the cavities and enables a high contrast for the homodyne detection interference. The fringe visibility
reaches 0.96. The shot noise level of all measurements is easily
obtained by blocking the output of the OPO. Let us emphasize that the pump is matched to the OPO cavity by temporarily inserting mirrors reflective at 426 nm and thus creating a cavity resonant for the blue pump. This solution turns out to be very efficient.

\begin{figure}[t!]
\begin{center}
\includegraphics[width=10cm]{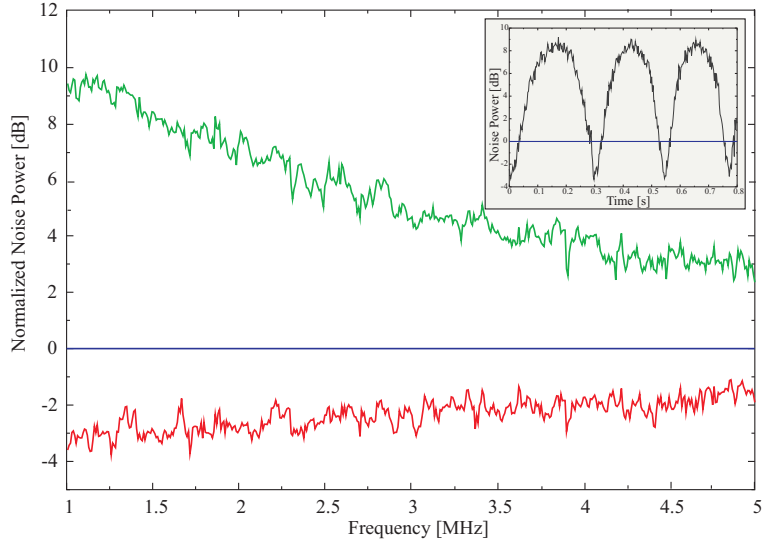}
\end{center}
\caption{Normalized noise variance for the squeezed and anti-squeezed quadratures, from 1 MHz to 5 MHz. The inset gives the noise variance at 1.5 MHz while scanning the phase of the local oscillator. The resolution bandwidth is set to 100 kHz and the video bandwidth to 100
Hz. 
}\label{broad}
\end{figure}

Figure \ref{broad} gives the noise variances of the squeezed and anti-squeezed quadratures for a frequency spectrum from 1 to 5 MHz.  The inset shows the noise variance while
scanning the local oscillator phase for a fixed noise analysis frequency of 1.5 MHz. For these measurements, the blue pump power was set to 75 mW. 3 dB of
squeezing is obtained, with an excess noise on the anti-squeezed quadrature around 9 dB.  This noise reduction value has to be compared to the
theoretical value $V$ given by \cite{fabre,kimble}
\begin{eqnarray} V=1-\frac{T}{T+L}\frac{4\sigma}{(1+\sigma)^2+4\Omega^2} 
\end{eqnarray}
where $T$ is the output coupler transmission, $L$ the additional intra-cavity losses due to absorption or scattering, $\Omega$ the analysis frequency normalized to the cavity bandwidth (10 MHz) and $\sigma$ the amplitude pump power normalized to the threshold. By taking
$\sigma=0.9$, $\Omega=0.1$, $T=0.07$ and $L=0.03$ (determined by measuring the cavity finesse and mirror transmissions) , the
expected value before detection produced at the OPO output is $-5$ dB. Let us note that $L$ is mostly due to absorption in PPKTP at this particular wavelength, as no pump-induced losses were measured.  The detector quantum
efficiency is estimated to be $0.90$, the fringe visibility is $0.96$
and the propagation efficiency is evaluated to be around $0.95$. These
values give an overall detection efficiency of $0.9 \cdot
0.96^{2} \cdot 0.95\simeq0.8$. After detection, the expected squeezing
is thus reduced to $-3.5$ dB, in good agreement with the experimental values.

\begin{figure}[t!]
\begin{center}
\includegraphics[width=9cm]{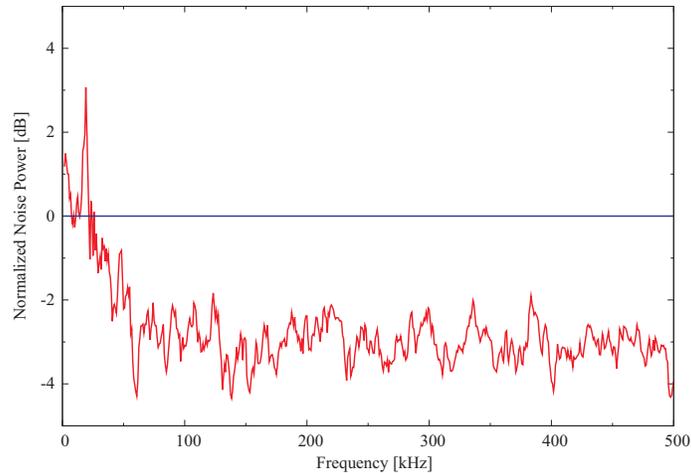}
\end{center}
\caption{Normalized noise variance up to 500
kHz after correction of the
electronic noise. The
resolution bandwidth is set to 30 kHz and the video bandwidth to 36
Hz.
}\label{bf}
\end{figure}

Figure \ref{bf} shows the broadband noise reduction similar to the Fig. \ref{broad} insert, but now for a lower frequency range,
between 0 and 500 kHz. Squeezing is expected to be higher in this range, but technical noise results in its degradation. The stability of the setup and noise of the laser are important parameters here. In particular, the lock beam power needs to be decreased as much as possible to avoid noise coupling into the device. In our setup, squeezing is finally detected down to 25 kHz, and 3$\pm$0.5 dB are observed for the 100-500 kHz frequency range. Measurements are corrected from the electronic dark noise. The presence of low-frequency sideband squeezing is a requisite for future quantum networking applications such as the storage of squeezed light by EIT, where the transparency window width is a limiting factor \cite{honda, appel}.

In conclusion, we have demonstrated the generation of squeezed light locked on the $D_2$ cesium line. More than 3 dB of noise reduction has been obtained and the squeezing is preserved for sideband frequencies down to 25 kHz. This ability opens the way to further investigations of light-matter interface using cesium atomic ensembles, like EIT or Raman storage of non-classical state of light in the continuous variable regime.

\section*{Acknowledgments}
This work was supported by the French ANR under the PNANO contract IRCOQ and by the EU under the projects COVAQIAL and COMPAS.
Xiaojun Jia acknowledges support from La Ville de
Paris. J. Ortalo acknowledges financial support for this work from the DGA
represented by B. Desruelle. We would like to thank J. Cviklinski for useful discussions in the early stage of the experiment.


\begin{thebibliography}{}

\bibitem{Schnabel} H. Vahlbruch, M. Mehmet, S. Chelkowski, B. Hage, A. Franzen, N. Lastzka, S. Go§ler, K. Danzmann, R. Schnabel,``Observation of Squeezed Light with 10-dB Quantum-Noise Reduction,'' Phys. Rev. Lett.  \textbf{100}, 033602  (2008).

\bibitem{zoller05} P. Zoller \textit{et al.}, ``Quantum information processing and communication, Strategic report on current status, visions and goals for research in Europe,'' Eur. Phys. J. D \textbf{36},
203-228 (2005).

\bibitem{cerf} N.J. Cerf, G. Leuchs, E.S. Polzik eds, Quantum Information with Continuous Variables, (World Scientific Publishing, New Jersey, 2007).

\bibitem{jeff} H.J. Kimble, ``The quantum internet,'' Nature \textbf{453},
1023-1030 (2008).

\bibitem{tanimura} T. Tanimura, D. Akamatsu, Y. Yokoi, A. Furusawa, M. Kozuma,``Generation of squeezed vacuum resonant on a rubidium $D_1$ line with periodically poled KTiOPO4,'' Opt. Lett. \textbf{31}, 2344-2346 (2006).


\bibitem{hetet} G. Hetet, O. Glockl, K.A. Pilypas, C.C. Harb, B.C. Buchler, H.A. Bachor, P.K. Lam,``Squeezed light for bandwith-limited atom optics experiments at the rubidium D1 line,'' J. Phys. B: At. Mol. Opt. Phys.  \textbf{40}, 221-226 (2007).

\bibitem{honda} K. Honda, D. Akamatsu, M. Arikawa, Y. Yokoi, K. Akiba, S. Nagatsuka, T. Tanimura, A. Furusawa, M. Kozuma, ``Storage and Retrieval of a Squeezed Vacuum," Phys. Rev. Lett.  \textbf{100},093601 (2008).

\bibitem{appel} J. Appel, E. Figueroa, D. Korystov, M. Lobino, A. I. Lvovsky,``Quantum memory for squeezed light,'' Phys. Rev. Lett.  \textbf{100}, 093602 (2008).


\bibitem{polzik} E.S. Polzik, J. Carri, H.J. Kimble,``Spectroscopy with squeezed light,'' Phys. Rev. Lett.  \textbf{68}, 3020  (1992).

\bibitem{marin} F. Marin, A. Bramati, V. Jost, E. Giacobino,``Demonstration of high sensitivity spectroscopy with squeezed semiconductor lasers,'' Optics Commun.  \textbf{140}, 146  (1997).

\bibitem{nielsen} J. S. Neergaard-Nielsen, B. Melholt Nielsen, C. Hettich, K. M¿lmer, E. S. Polzik,``Generation of a Superposition of Odd Photon Number States for Quantum Information Networks,'' Phys. Rev. Lett.  \textbf{97}, 083604  (2006).

\bibitem{LKB1} F. Villa, A. Chiummo, E. Giacobino, A. Bramati ,``High-efficiency blue-light generation with a ring cavity with periodically poled KTP," J. Opt. Soc. Am. B  \textbf{24},
576-580 (2007).

\bibitem{tilt} D.A. Shaddock, M.B. Gray, D.E. McClelland ,``Frequency locking a laser to an optical cavity by use 
of spatial mode interference," Opt. Lett.  \textbf{24}, 1499
 (1999).


\bibitem{PDH} R. W. P. Drever, J. L. Hall, F. V. Kowalski, J. Hough, G. M. Ford, A. J. Munley and H. Ward ,``Laser phase and frequency stabilization using an optical resonator,'' Appl. Phys. B  \textbf{31}, 97
 (1983).


\bibitem{fabre} C. Fabre, S. Reynaud , in Fundamental Systems in Quantum Optics, Les Houches 1990, J. Dalibard, J. M. Raimond, J. Zinn-Justin, Eds. (Elsevier, Amsterdam, 1992)
\bibitem{kimble} H.J. Kimble, in Fundamental Systems in Quantum Optics, Les Houches 1990, J. Dalibard, J. M. Raimond, J. Zinn-Justin, Eds. (Elsevier, Amsterdam, 1992)


  
\end{thebibliography}
\end{document}